\documentclass[conference]{IEEEtran}
\IEEEoverridecommandlockouts

\usepackage{cite}
\usepackage{amsmath,amssymb,amsfonts}
\usepackage{algorithmic}
\usepackage{graphicx}
\usepackage{textcomp}
\usepackage{xcolor}
\def\BibTeX{{\rm B\kern-.05em{\sc i\kern-.025em b}\kern-.08em
    T\kern-.1667em\lower.7ex\hbox{E}\kern-.125emX}}

\begin{document}

\title{Auditory Attention Decoding without \\Spatial Information: A Diotic EEG Study}

\author{
\IEEEauthorblockN{Masahiro Yoshino\IEEEauthorrefmark{1},
Haruki Yokota\IEEEauthorrefmark{2},
Junya Hara\IEEEauthorrefmark{2},
Yuichi Tanaka\IEEEauthorrefmark{2},
Hiroshi Higashi\IEEEauthorrefmark{2}}
\IEEEauthorblockA{\IEEEauthorrefmark{1}Department of Electronic and Information
Engineering, School of Engineering, The University of Osaka, Suita, Japan\\
\IEEEauthorrefmark{2}Graduate School of Engineering, The University of Osaka, Suita, Japan \\
Email: m.yoshino@msp-lab.org}
}

\maketitle

\begin{abstract}
Auditory attention decoding (AAD) identifies the attended speech stream in multi-speaker environments by decoding brain signals such as electroencephalography (EEG). 
This technology is essential for realizing smart hearing aids that address the cocktail party problem and for facilitating objective audiometry systems.
Existing AAD research mainly utilizes \textit{dichotic} environments where different speech signals are presented to the left and right ears, 
enabling models to classify directional attention rather than speech content. 
However, this spatial reliance limits applicability to real-world scenarios, such as the ``cocktail party'' situation, where speakers overlap or move dynamically.
To address this challenge, we propose an AAD framework for \textit{diotic} environments where identical speech mixtures are presented to both ears, eliminating spatial cues. 
Our approach maps EEG and speech signals into a shared latent space using independent encoders. 
We extract speech features using wav2vec 2.0 and encode them with a 2-layer 1D convolutional neural network (CNN), while employing the BrainNetwork architecture for EEG encoding. 
The model identifies the attended speech by calculating the cosine similarity between EEG and speech representations.
We evaluate our method on a diotic EEG dataset and achieve 72.70\% accuracy, which is 22.58\% higher than the state-of-the-art direction-based AAD method.
\end{abstract}


\section{Introduction}
Auditory attention decoding (AAD)~\cite{b7,b8,b9} aims to identify the attended speech stream 
in multi-speaker environments by decoding brain signals 
such as electroencephalography (EEG)~\cite{b1}.
This technology is essential for realizing smart hearing aids~\cite{b3}, capable of selectively enhancing a specific speech source in noisy environments, 
thereby addressing the so-called cocktail party problem~\cite{b6}. Furthermore, AAD is expected to improve understanding of neural mechanisms of selective listening 
and facilitates the development of objective audiometry systems~\cite{b4, b5}.

Existing AAD research is primarily classified into two approaches based on the decoding strategy. The first approach detects the attended speech by analyzing 
the correspondence between EEG and speech signals~\cite{b30, b31}.
The second approach classifies the attended direction of the speech based on only the EEG signals~\cite{b32, b14, b33, b34}. DARNet~\cite{b14}, a state-of-the-art deep learning model, adopts 
the directional strategy and achieves 96.2\% accuracy on the standard KUL dataset~\cite{b29}.

These AAD approaches have supporsed and evaluated with \textit{dichotic} environments.
In this environment, speech signals are clearly separated into the left and right ears~\cite{b35, b36, b37}.
This separation makes it easier to classify based on spatial cues, which effectively serve as explicit directional information to distinguish the target speaker.
The AAD models can capture this directional information and decode the direction of the target speaker. However, these spatial cue-based models may not work in real-world cocktail party scenarios where speakers often stand close together or move dynamically.
Consequently, methods using spatial cues may have limited applicability. 

To build AAD systems that identify the attended speech even when sound sources overlap or move, we need to consider a \textit{diotic} scenario. In this scenario, the exact same mixture of speech streams is simultaneously presented to both ears.
The spatial cues are identical across streams and provide no discriminative information for source localization.

In this paper, to address the AAD problem in the diotic setting, 
we propose a framework based on speech features.
Our approach maps EEG and speech signals into 
a shared latent space using independent encoders.
Specifically, we extract speech features using a pre-trained wav2vec 2.0 model~\cite{b17} 
and encode them using a 2-layer 1D convolutional neural network (CNN). Simultaneously, 
we employ the BrainNetwork architecture~\cite{b16} to encode EEG signals.
In the shared latent space, we calculate the cosine similarity 
between the EEG and each speech stream.
By evaluating this direct correspondence, 
our model identifies the attended speech based on speech content without using spatial information.

We evaluate the proposed method using the diotic EEG dataset in~\cite{b18}.
Our method achieves 72.70\% accuracy on five-second segments.
In contrast, DARNet performs 50.12\% in the diotic setting, which is close to chance level.
This result validates that our proposed framework effectively solves the diotic AAD problem based on speech content without relying on spatial cues.

Furthermore, we validate the neuroscientific plausibility of our model 
through two analyses. Based on the \textit{Late Selection Theory}~\cite{b19}, 
we validate that our model distinguishes speeches at the attentional selection stage 
rather than the early acoustic processing stage.
Additionally, using SHAP analysis~\cite{b20} in the context of \textit{Top-down Gain Control}~\cite{b26}, 
we show that our model specifically highlights EEG channels associated with attentional control. 
These analyses suggest that our framework captures the neural mechanisms of selective speech attention.

\section{Diotic Auditory Attention Decoding}
This section presents our proposed diotic AAD framework designed to overcome the spatial reliance of conventional AAD methods in dichotic settings.
\subsection{Task}
\begin{figure}[t]
\centering
\includegraphics[width=0.95\linewidth]{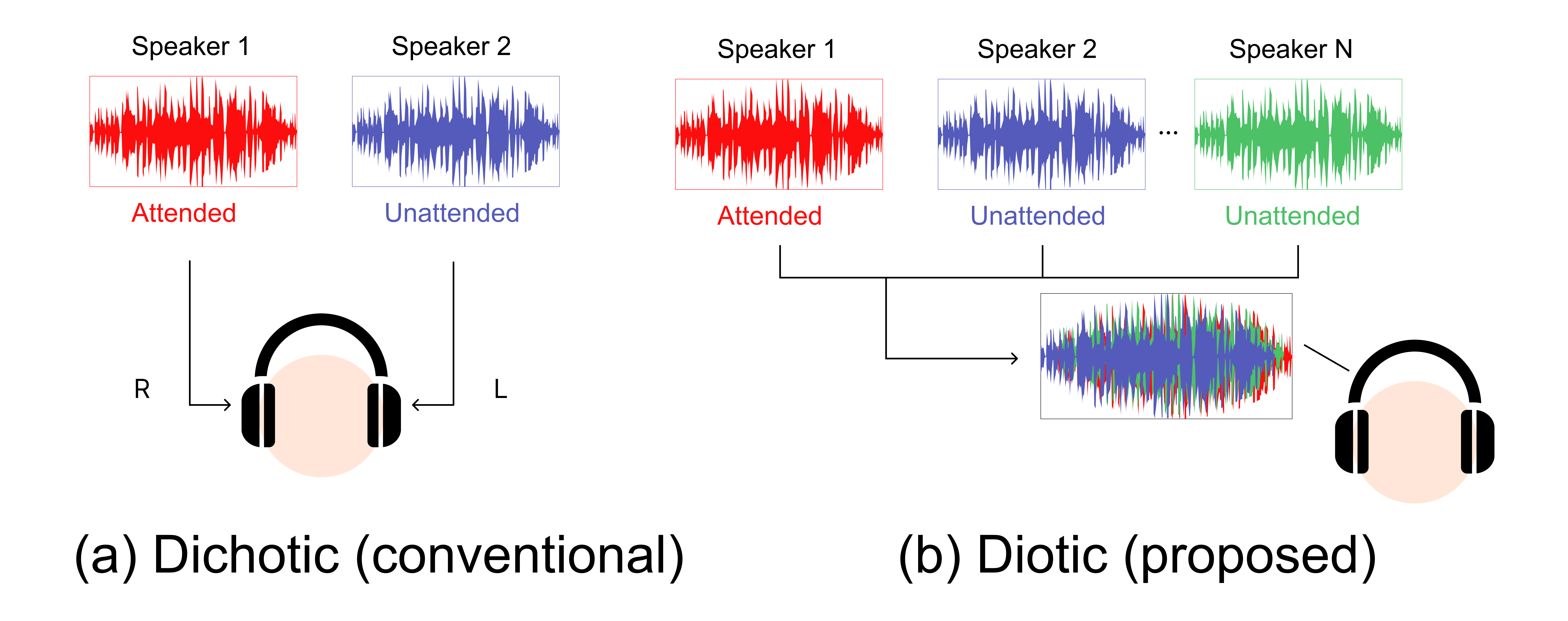}
\caption{Comparison between the dichotic and diotic settings. (a) The dichotic setup presents distinct speech streams to each ear that enables decoding based on spatial location. (b) The diotic setup presents a mixture of $N$ speech streams to both ears. This requires the decoder to identify the attended speech without using speech source direction information.}
\label{fig:diotic_concept}
\end{figure}

Figure~\ref{fig:diotic_concept} illustrates the concept of the diotic AAD task along with the dichotic setting. The goal of the diotic AAD task is to identify the attended speech stream from a set of multiple speech streams.
Unlike dichotic environments where different speech signals are presented to each ear, the diotic setup presents a mixture of $N$ speech streams to both ears simultaneously.

Let $\mathbf{E} \in \mathbb{R}^{C \times T}$ denote an EEG segment consisting of $C$ channels and $T$ time instances, and let $\{\mathbf{S}_1, \dots, \mathbf{S}_N\}$ be a set of $N$ candidate speech streams.
Here, each stream $\mathbf{S}_i \in \mathbb{R}^{F \times T}$ has a feature dimension $F$.
Given these inputs, the task is to predict the target index $\hat{y} \in \{1, \dots, N\}$ corresponding to the attended speech.
Using a classifier function $g(\cdot ; \boldsymbol{\phi})$ parameterized by learnable weights $\boldsymbol{\phi}$, the AAD task is formulated as
\begin{equation}
    \hat{y} = g(\mathbf{E}, \mathbf{S}_1, \mathbf{S}_2, \dots, \mathbf{S}_N; \boldsymbol{\phi}).
    \label{eq:diotic_task}
\end{equation}

\subsection{Framework \label{ssec:framework}}
\begin{figure}[t]
\centering
\includegraphics[width=0.95\linewidth]{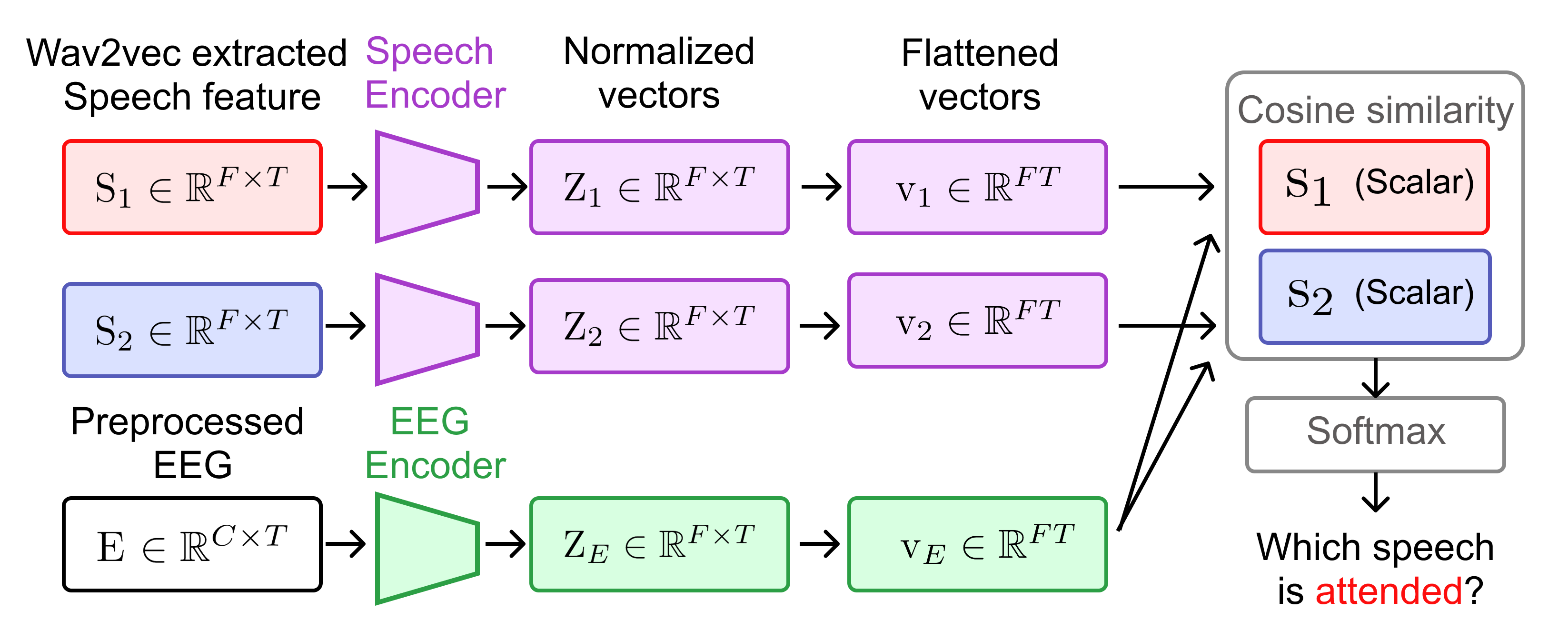}
\caption{Overview of the proposed diotic AAD framework.}
\label{fig:framework}
\end{figure}

Figure~\ref{fig:framework} shows the overview of our proposed diotic AAD framework.
The framework consists of three main stages: feature extraction, similarity calculation, and training.

\subsubsection{Feature Extraction}
The model maps a preprocessed EEG segment $\mathbf{E}$ and candidate speech streams $\mathbf{S}_i$ into a shared latent space using independent encoders.
The detailed preprocessing procedures for EEG and speech signals are described in Sections~\ref{ssec:eeg_preprocessing} and~\ref{ssec:speech_preprocessing}, respectively.

For the speech streams, we utilize wav2vec 2.0 features as inputs, which are known to improve the detection of synchronization between EEG and audio signals~\cite{b16}.
These features are transformed by a speech encoder consisting of a 2-layer 1D convolutional neural network (CNN).

For the EEG inputs, we employ the BrainNetwork architecture~\cite{b16}, which consists of a spatial attention layer and a 5-block residual network (ResNet).

As a result, we obtain feature maps $\mathbf{Z}_{E}$ for the EEG signals and $\mathbf{Z}_{i}$ for the $i$th speech stream from the respective encoders, where $\mathbf{Z}_{l} \in \mathbb{R}^{F \times T}$ for $l \in \{E, 1, \dots, N\}$, 
$F$ denotes the feature dimension, and $T$ denotes the time length.

\subsubsection{Similarity Calculation}
We apply a learnable weight vector $\mathbf{w} \in \mathbb{R}^{F}$ to the feature maps via channel-wise scaling.
This allows the model to emphasize important features in the latent representation.
The weighted representation $\tilde{\mathbf{Z}}$ is calculated as
For any feature map $\mathbf{Z} \in \{\mathbf{Z}_E, \mathbf{Z}_i\}$, the weighted representation $\tilde{\mathbf{Z}}$ is calculated as
\begin{equation}
    \tilde{\mathbf{Z}} = \operatorname{diag}(\mathbf{w}) \mathbf{Z},
\end{equation}
where $\operatorname{diag}(\mathbf{w})$ denotes a diagonal matrix with the elements of $\mathbf{w}$.
We then flatten the weighted feature maps $\tilde{\mathbf{Z}}_E$ and $\tilde{\mathbf{Z}}_i$ into one-dimensional vectors $\mathbf{v}_E, \mathbf{v}_i \in \mathbb{R}^{FT}$.
Finally, the similarity score $s_i$ for the $i$th candidate speech stream is calculated using cosine similarity as
\begin{equation}
    s_{i} = \frac{\mathbf{v}_{E}^\top \mathbf{v}_{i}}{\|\mathbf{v}_{E}\| \|\mathbf{v}_{i}\|}.
\end{equation}

\subsubsection{Training}
In this study, we focus on the binary classification scenario ($N=2$).
This setting aligns with the experimental design of the dataset and allows for direct comparison with dichotic AAD methods.

A training sample is defined as $\{\mathbf{E}, \mathbf{S}_1, \mathbf{S}_2, y\}$, where $\mathbf{S}_1$ and $\mathbf{S}_2$ are the candidate speech streams and $y \in \{1, 2\}$ is the index of the attended stream.
To predict the target index, the model calculates the probability $\hat{p}_i$ from the similarity scores $s_i$ ($i \in \{1, 2\}$) using the softmax function with a temperature parameter $\tau$ as
\begin{equation}
\hat{p}_i = \frac{\exp(s_{i} / \tau)}{\sum_{j=1}^{2} \exp(s_{j} / \tau)}.
\end{equation}
During the training phase, to maximize the likelihood of the correct stream $\hat{p}_y$, we use the cross-entropy loss function defined as
\begin{equation}
\mathcal{L} = - \log(\hat{p}_y).
\end{equation}

\section{Experiments and Results}
We conduct AAD experiments using NVIDIA GeForce RTX 3090 24GB GPUs with Driver Version 570.195.03, CUDA Version 12.8, and Python 3.11.12.

\subsection{Dataset}
We employ a publicly available diotic listening dataset from Stoll et al.~\cite{b18}. The experiment1 diotic subset contains EEG data from 28 subjects who listen binaurally to simultaneous speech from two speakers and are instructed to attend to one speaker. 
Audio stimuli are from male and female speakers. Each subject completes 120 listening sessions of 64~seconds each. For experimental consistency, the female voice pitch is lowered to match the male voice frequency. EEG data were recorded with 32 channels at a 10~kHz sampling rate. 

\subsection{EEG Preprocessing \label{ssec:eeg_preprocessing}}
The EEG data processing pipeline is implemented using the MNE-Python library.
First, we convert the signal units from Volts (V) to microvolts ($\mu$V) to ensure numerical stability for the deep learning model.
Next, we apply a Finite Impulse Response (FIR) bandpass filter between 0.5~Hz and 32.0~Hz to remove DC drift and high-frequency noise.
After filtering, we re-referenced the signals using the Common Average Reference (CAR) method~\cite{b22}, which subtracts the average potential of all 32 channels from each channel.
These frequency bands are known to be relevant to auditory processing~\cite{b21}. Finally, the data were downsampled to 64~Hz.

\subsection{Speech Preprocessing \label{ssec:speech_preprocessing}}
We follow the feature extraction pipeline used in~\cite{b16}. We use the self-supervised speech model wav2vec 2.0~\cite{b17} to extract speech representations. In practice, we use the wav2vec2-large-960h. 
We utilize the output of the model's 14th layer as the speech representation. The representation initially has a dimensionality of 1024, which we then reduce to 64 dimensions using Principal Component Analysis (PCA).
To match the sampling frequency of the EEG data, we resample the wav2vec 2.0 speech representations to 64~Hz.

\subsection{Decoding Procedure}
We compare two models: DARNet and the proposed model for solving diotic AAD. We set segment length to 1, 3 and 5~seconds.
To prevent data leakage, we divide the 28 subjects into training (20), validation (4), and test (4) sets. We employ a 7-fold cross-validation for a reliable assessment of the model's generalization capability.

\subsubsection{DARNet}
We train the model using the Adam optimizer~\cite{b23} with learning rate of $5 \times 10^{-4}$, weight decay of $3 \times 10^{-4}$, and batch size of 32. We conduct training for a maximum of 100~epochs with early stopping patience of 10~epochs. These hyperparameters follow the original DARNet configuration~\cite{b14}.

\subsubsection{Our Model}
We follow the architecture defined in Section~\ref{ssec:framework}. We train the model using the Adam optimizer~\cite{b23} with a learning rate of $2 \times 10^{-5}$ and a batch size of 32 for a maximum of 50~epochs. We apply early stopping with a patience of 20~epochs based on validation loss.
The softmax temperature parameter is set to $\tau=0.05$ (corresponding to a scaling factor of 20).

\begin{figure}[t]
\centering
\includegraphics[width=0.95\linewidth]{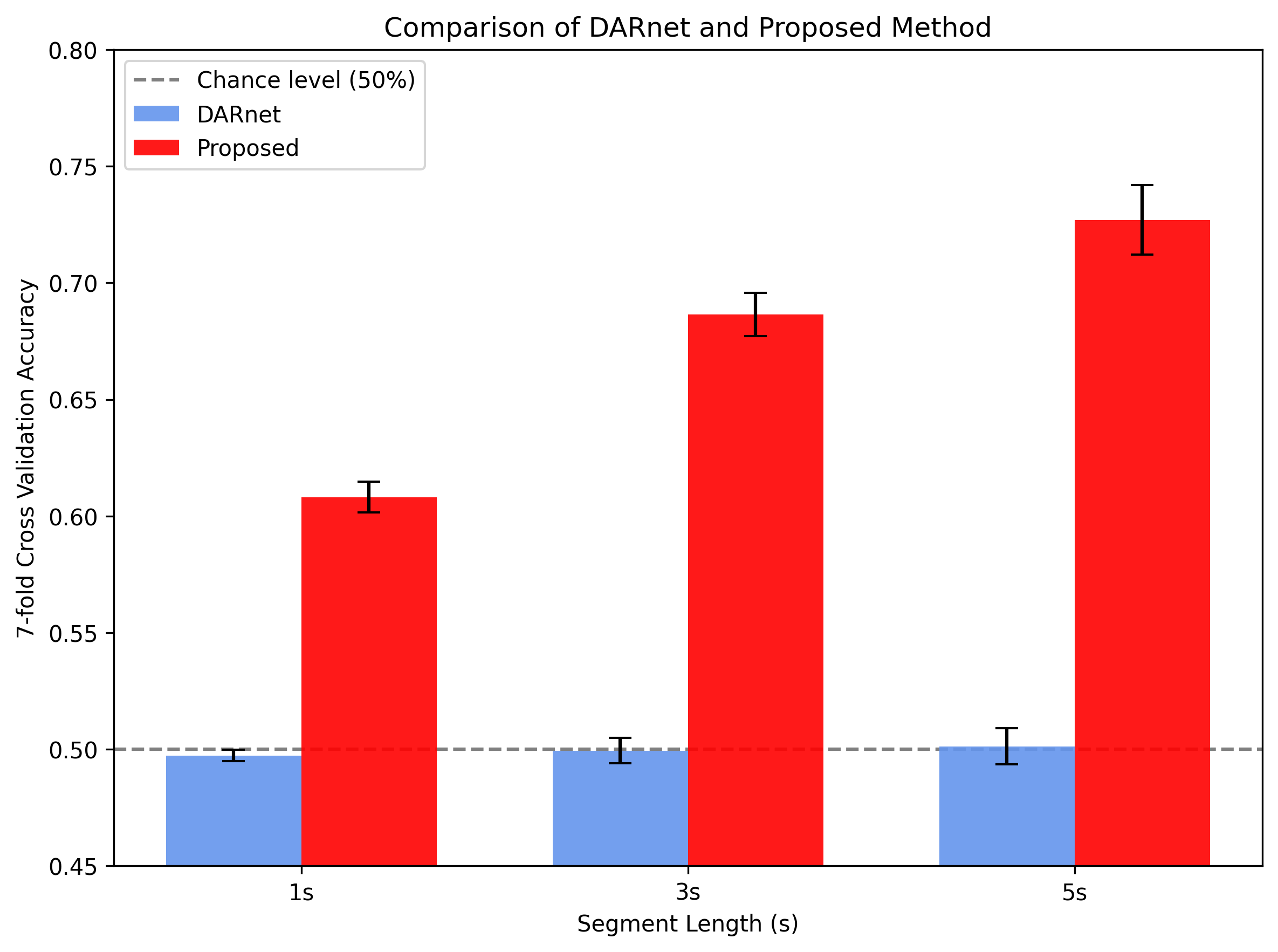}
\caption{Comparison of classification accuracy between DARNet and the proposed method across different segment lengths (1s, 3s, and 5s). The bar heights represent the mean accuracy obtained from 7-fold cross-validation, and the error bars indicate the standard deviation. The dashed gray line represents the chance level (50\%).}
\label{fig:main_results}
\end{figure}

\subsection{Results}
We compare the two architectures' performance on the diotic AAD task to determine whether attention decoding without speech source direction information is possible. 
Figure~\ref{fig:main_results} summarizes the results of the diotic AAD task. DARNet achieves chance-level accuracy, indicating DARNet's reliance on speech source direction, which requires dichotic presentation for this model. 
In contrast, our framework achieves 60.80\% (1~second), 68.65\% (3~seconds), and 72.70\% (5~seconds) accuracy, respectively, significantly exceeding chance level. This demonstrates that the proposed model enables successful attention decoding without using speech source direction information.

\section{Discussion}
While our results demonstrate the feasibility of diotic AAD based on EEG-speech correspondence, 
for practical applications such as smart hearing aids, it is crucial to verify whether this performance stems from attentional selection.
To validate this, we conduct two further analyses based on cognitive neuroscience theories: the \textit{Late Selection Theory}~\cite{b19} 
and the \textit{Top-down Gain Control} mechanism~\cite{b26}.

For these analyses, we introduce the Match-Mismatch (M-MM) task~\cite{b10} 
as a control condition. In the M-MM task, the model classifies 
whether a single speech stream temporally matches the recorded EEG.
Unlike the AAD task, the M-MM task measures the correspondence between EEG and speech without attentional influence.
Therefore, this task reflects fundamental acoustic processing independent of the selective attention state.

\subsection{Analysis Based on the Late Selection Theory}
Based on the Late Selection Theory~\cite{b19}, 
auditory processing consists of two stages: 
an early acoustic processing stage where all sounds are processed, 
and a later attentional selection stage.
Since the M-MM task is attention-independent, it can be interpreted 
as decoding the neural activity of the early acoustic processing stage.
We hypothesize that if our model correctly extracts 
these pre-selection features, it should be able to detect temporal alignment 
for both attended and unattended streams equally well.
Conversely, if a significant performance gap occurs, 
it would suggest that the model is based on early sensory filtering, 
contradicting the premise of late-stage attention decoding.

Formally, let $\mathbf{E}_{t} \in \mathbb{R}^{C \times T}$ denote 
the EEG segment at time $t$. We construct a training sample $\{\mathbf{E}_{t}, \mathbf{S}_1, \mathbf{S}_2, y\}$,
where $\{\mathbf{S}_1, \mathbf{S}_2\}$ contains one temporally matching speech segment to the EEG signal (a speech segment presented at time $t$) and one temporally mismatched segment (a speech segment presented at a different time $t'$).
The label $y \in \{1, 2\}$ indicates the index of the matching segment.
The task is formulated as
\begin{equation}
    \hat{y} = g(\mathbf{E}_{t}, \mathbf{S}_1, \mathbf{S}_2; \boldsymbol{\phi}).
\end{equation}
The function $g(\cdot; \boldsymbol{\phi})$ is a classifier parameterized by learnable weights $\boldsymbol{\phi}$.

\begin{figure}[t]
\centering
\includegraphics[width=0.95\linewidth]{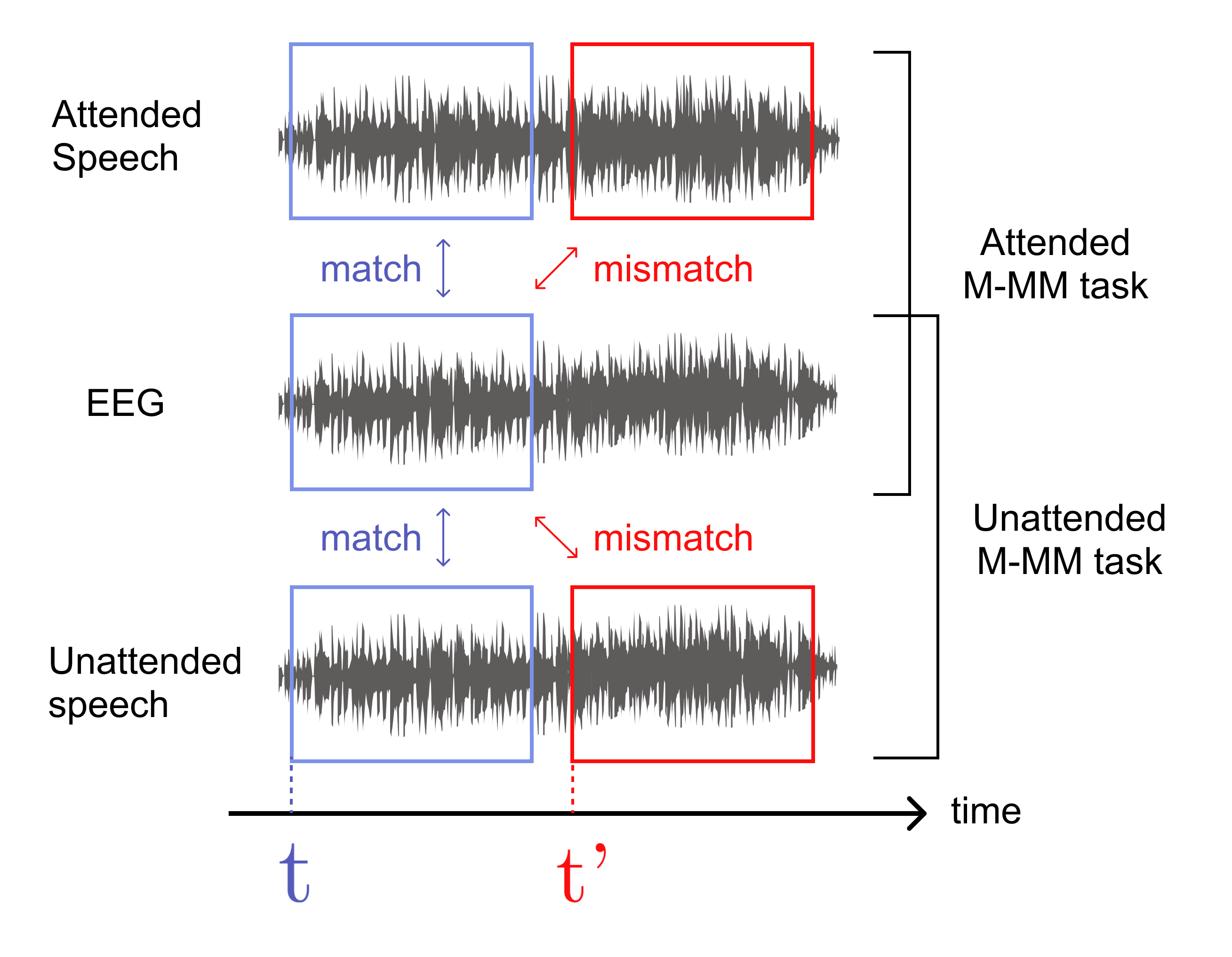}
\caption{Conceptual diagram of the M-MM task. The goal is to verify whether our proposed framework operates through solely acoustic matching or late-stage attention by comparing acoustic decoding accuracy between attended and unattended streams. EEG segment at time $t$ is paired with temporally matching speech segments and temporally mismatching segments (at different time $t'$). The model discriminates which speech segment temporally corresponds to the EEG, independently for attended and unattended streams, to evaluate speech decoding accuracy independent of attentional state.}
\label{fig:diotic_mmm_concept}
\end{figure}

We train and evaluate two separate models for this M-MM task using both attended and unattended streams. Both models use the same architecture described in Section~\ref{ssec:framework}. The models are independently trained and evaluated using datasets consisting of attended and unattended streams, respectively.
Figure~\ref{fig:diotic_mmm_concept} illustrates this experimental design.
Table~\ref{tab:mm_results} shows the results.
We confirm that the models achieve comparable accuracy for both attended (84.37\%) and unattended (83.30\%) streams.
This similarity demonstrates that the brain captures acoustic features even from unattended sources, implying that the acoustic processing stage does not differentiate between the streams.
Consequently, the high accuracy of our diotic AAD model—which successfully distinguishes the attended stream—suggests that our method operates at the subsequent attentional selection stage, utilizing features that emerge after this early acoustic processing.

\begin{table}[t]
\caption{M-MM Results (\%)}
\begin{center}
\begin{tabular}{|l|c|}
\hline
\textbf{Stream Type} & \textbf{Accuracy (\%)} \\
\hline
chance level & 50.0 \\
Attended Stream & 84.37 $\pm$ 1.02 \\
Unattended Stream & 83.30 $\pm$ 1.21 \\
\hline
\end{tabular}
\label{tab:mm_results}
\end{center}
\end{table}

\subsection{Neural Mechanisms: SHAP Explainability Analysis}
Second, we investigate the neural basis of the model's decision-making to demonstrate that the model leverages specific brain regions associated with attention.
The Top-down Gain Control theory~\cite{b26} suggests that attentional selection involves prefrontal regions modulating the auditory cortex to enhance the representation of the target stream.
To support this, we employ SHAP (SHapley Additive exPlanations)~\cite{b20} to compare the important EEG features in the AAD task with those in the M-MM task.
By contrasting the AAD task (which requires selection) with the M-MM task (which reflects basic acoustic processing), we aim to isolate the specific neural correlates of attention.

\begin{figure}[t]
\centering    
\includegraphics[width=0.95\linewidth]{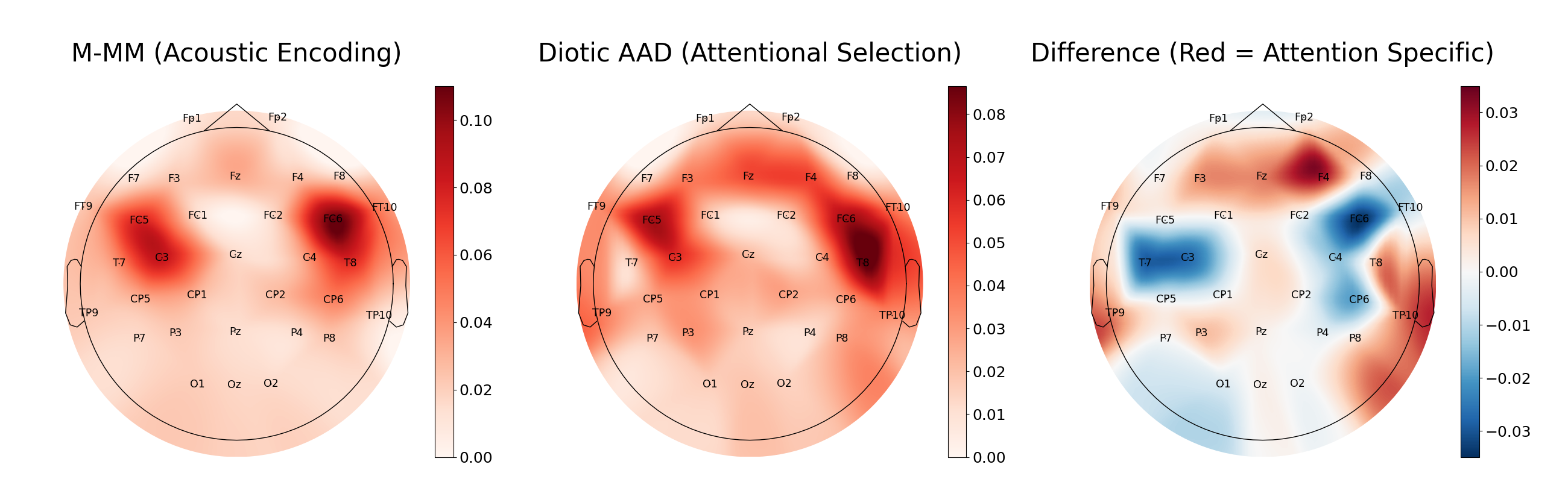}
\caption{Comparison of SHAP-based EEG spatial importance. 
Left: M-MM task (acoustic encoding) exhibits importance clusters in central-lateral regions (FC6, C3, FC5, T8). This reflects auditory sensory processing. 
Middle: Diotic AAD task (attentional selection) shows distinct importance of frontal (Fz, F3, F4) and temporo-parietal (TP9, TP10) regions. 
Right: The difference map highlights attention-specific frontal engagement and a relative decrease in central-lateral importance. This is consistent with the recruitment of top-down control networks.}
\label{fig:shap}
\end{figure}

We generate spatial importance maps for both tasks using GradientExplainer and calculate the difference map.
Figure~\ref{fig:shap} shows the EEG spatial importance of the models for each task.
In the M-MM task (Left), feature importance concentrates in temporal and fronto-central regions (e.g., T8, FC6, C3).
This suggests that the model prioritizes early acoustic and temporal processing common to both tasks.
In contrast, the diotic AAD task (Middle) highlights distinct frontal regions (Fz, F3, F4) and temporo-parietal regions (TP9, TP10).
The prominence of these frontal areas is consistent with the top-down gain control mechanism.
These results indicate that our model shows increased contribution from EEG channels associated with frontal attentional control during the AAD task.
This functional shift supports the interpretation that our proposed framework goes beyond simple acoustic matching and reflects the neural mechanisms of selective attention.

\subsection{Limitations}
Our study has several limitations.

\subsubsection{Limited Dataset Verification}
Our evaluation is currently limited to a single dataset.
This constraint arises because, to the best of our knowledge, the dataset utilized in this study is the only publicly available dataset containing EEG signals acquired during selective attention tasks in a diotic environment.

\subsubsection{Influence of Acoustic Cues}
Although we utilize pitch-equalized male and female voices, residual acoustic cues (e.g., timbre and formant structure) may still facilitate segregation.
This potentially reduces the ecological validity of the strict top-down attention task.

\subsubsection{Model Interpretability}
While SHAP analysis confirms spatial consistency with known auditory cortical activity, this result reflects the feature attribution of the deep learning model rather than a direct measurement of neural processing.

\subsubsection{Scalability to Multiple Speakers}
As formulated in Equation~\ref{eq:diotic_task}, the diotic AAD task is theoretically applicable to scenarios where three or more speech streams are presented.
However, the current study is limited to the binary classification scenario ($N=2$).
Validating the model's performance in such complex multi-speaker environments requires a dataset containing diotic EEG recordings with three or more speakers.

\section{Conclusion}
We address a fundamental challenge in auditory attention research: how to realize attention-based speech selection without speech source direction information. 
We introduce the diotic AAD approach, which eliminates speech source direction information by presenting the same mixed speech to both ears. 
We evaluate two architectures and obtain contrasting results: DARNet (50.12\%) performs at chance level due to spatial reliance on speech source direction, 
while our proposed framework achieves 72.70\% accuracy. To explain this success, we conduct analyses using M-MM task and SHAP explainability. This shows that our framework effectively captures attention-based speech selection.

\section*{Acknowledgment}
This work was supported by JSPS KAKENHI Grant Numbers JP23H01415, JP23K26110, JP23K17461, 22H05163, and JP24K15047.

\section*{Code Availability}
The code to implement our work will be available on Github: https://github.com/yossiiiii-m/DioticAAD

\bibliographystyle{IEEEtran}
\bibliography{ref}

@article{b1,
  author    = {Inyong Choi and Siddharth Rajaram and Lenny A. Varghese and Barbara G. Shinn-Cunningham},
  title     = {Quantifying attentional modulation of auditory-evoked cortical responses from single-trial electroencephalography},
  journal   = {Frontiers in Human Neuroscience},
  volume    = {7},
  pages     = {115},
  year      = {2013}
}

@article{b3,
  author    = {Simon Geirnaert and Tom Francart and Alexander Bertrand},
  title     = {Electro\-encephalography-based auditory attention decoding: Toward neurosteered hearing devices},
  journal   = {IEEE Signal Processing Magazine},
  volume    = {38},
  number    = {4},
  pages     = {89--102},
  year      = {2021}
}

@article{b4,
  author    = {N. Sriraam},
  title     = {{EEG} based automated detection of auditory loss: A pilot study},
  journal   = {Expert Systems with Applications},
  volume    = {39},
  pages     = {723--731},
  year      = {2012}
}

@article{b5,
  author    = {M. Prabhu Prasad and B. Ram Gopal and Ch. Srinivasa Rao and A. Anjaneyulu},
  title     = {{EEG} Based Detection of Conductive and Sensorineural Hearing Loss using Artificial Neural Networks},
  journal   = {Journal of Next Generation Information Technology},
  volume    = {4},
  pages     = {204--212},
  year      = {2013}
}

@article{b6,
  author    = {E. C. Cherry},
  title     = {Some experiments on the recognition of speech, with one and with two ears},
  journal   = {The Journal of the Acoustical Society of America},
  volume    = {25},
  number    = {5},
  pages     = {975--979},
  year      = {1953}
}

@article{b7,
  author    = {Cong Han and James O'Sullivan and Yi Luo and Jose Herrero and Ashesh D. Mehta and Nima Mesgarani},
  title     = {Speaker-independent auditory attention decoding without access to clean speech sources},
  journal   = {Science Advances},
  volume    = {5},
  number    = {5},
  pages     = {eaav6134},
  year      = {2019}
}

@inproceedings{b8,
  author    = {Mohammad Jalilpour Monesi and Bernd Accou and Jair Montoya-Martinez and Tom Francart and Hugo Van Hamme},
  title     = {An {LSTM} based architecture to relate speech stimulus to {EEG}},
  booktitle = {2020 IEEE International Conference on Acoustics, Speech and Signal Processing (ICASSP)},
  pages     = {941--945},
  year      = {2020}
}

@article{b9,
  author    = {Siqi Cai and Enze Su and Longhan Xie and Haizhou Li},
  title     = {{EEG}-based auditory attention detection via frequency and channel neural attention},
  journal   = {IEEE Transactions on Human-Machine Systems},
  volume    = {52},
  number    = {2},
  pages     = {256--266},
  year      = {2021}
}

@article{b10,
  author    = {Bernd Accou and Mohammad Jalilpour Monesi and Hugo Van hamme and Tom Francart},
  title     = {Modeling the relationship between acoustic stimulus and {EEG} with a match-mismatch task},
  journal   = {Journal of Neural Engineering},
  volume    = {18},
  number    = {4},
  pages     = {046040},
  year      = {2021}
}

@inproceedings{b14,
  author    = {Haoyu Cheng and Huan Liao and Haolin Zhu and Pengyuan Zhang and Xin Xu},
  title     = {{DARnet}: Deep attention recurrent network for auditory attention detection},
  booktitle = {Proc. IEEE Int. Conf. Acoust., Speech, Signal Process. (ICASSP)},
  pages     = {8671--8675},
  year      = {2024}
}

@inproceedings{b16,
  author    = {Jiawei Wang and Yang Xiao and Tao Gu and Wei Xiang and Dingqi Yang},
  title     = {Self-supervised speech representation and {EEG} encoder for improved auditory attention decoding},
  booktitle = {Proc. IEEE Int. Conf. Acoust., Speech, Signal Process. (ICASSP)},
  pages     = {1856--1860},
  year      = {2024}
}

@inproceedings{b17,
  author    = {Alexei Baevski and Yuhao Zhou and Abdelrahman Mohamed and Michael Auli},
  title     = {wav2vec 2.0: A framework for self-supervised learning of speech representations},
  booktitle = {Advances in Neural Information Processing Systems (NeurIPS)},
  year      = {2020},
  pages     = {12449--12460}
}

@article{b18,
  author    = {Jonas Stoll and Alejandro Ojeda and Edmund C. Lalor and Piotr Majdak},
  title     = {The auditory brainstem response to natural speech is not affected by selective attention},
  journal   = {bioRxiv},
  year      = {2025}
}

@article{b19,
  author    = {J. A. Deutsch and D. Deutsch},
  title     = {Attention: Some theoretical considerations},
  journal   = {Psychological Review},
  volume    = {70},
  number    = {1},
  pages     = {80--90},
  year      = {1963}
}

@inproceedings{b20,
  author    = {Scott M. Lundberg and Su-In Lee},
  title     = {A unified approach to interpreting model predictions},
  booktitle = {Advances in Neural Information Processing Systems (NeurIPS)},
  year      = {2017},
  pages     = {4765--4774}
}

@article{b21,
  author    = {Nai Ding and Jonathan Z. Simon},
  title     = {Emergence of neural encoding of auditory objects while listening to competing speakers},
  journal   = {Proceedings of the National Academy of Sciences},
  volume    = {109},
  number    = {29},
  pages     = {11854--11859},
  year      = {2012}
}

@article{b22,
  author    = {D. J. McFarland and L. M. McCane and S. V. David and J. R. Wolpaw},
  title     = {Spatial filter selection for {EEG}-based communication},
  journal   = {Electroencephalography and Clinical Neurophysiology},
  volume    = {103},
  number    = {3},
  pages     = {386--394},
  year      = {1997}
}

@inproceedings{b23,
  author    = {Diederik P. Kingma and Jimmy Ba},
  title     = {{Adam}: A method for stochastic optimization},
  booktitle = {International Conference on Learning Representations (ICLR)},
  year      = {2015}
}

@article{b26,
  author    = {J. B. Fritz and S. Shamma and M. Elhilali and D. Klein},
  title     = {Auditory attention---focusing the searchlight on sound},
  journal   = {Current Opinion in Neurobiology},
  volume    = {17},
  number    = {4},
  pages     = {437--455},
  year      = {2007}
}

@article{b29,
  title = {Auditory Attention Detection Dataset Kuleuven},
  author = {Das, Neetha and Francart, Tom and Bertrand, Alexander},
  date = {2019-08},
  publisher = {Zenodo},
  version = {2.0}
}

@article{b30,
  author={Aroudi, Ali and Marquardt, Daniel and Daclo, Simon},
  booktitle={2018 IEEE International Conference on Acoustics, Speech and Signal Processing (ICASSP)}, 
  title={EEG-Based Auditory Attention Decoding Using Steerable Binaural Superdirective Beamformer}, 
  year={2018},
  volume={},
  number={},
  pages={851-855},
}

@article{b31,
  author={Das, Neetha and Van Eyndhoven, Simon and Francart, Tom and Bertrand, Alexander},
  booktitle={2017 25th European Signal Processing Conference (EUSIPCO)}, 
  title={EEG-based attention-driven speech enhancement for noisy speech mixtures using N-fold multi-channel Wiener filters}, 
  year={2017},
  volume={},
  number={},
  pages={1660-1664},
}

@ARTICLE{b32,
  author={Su, Enze and Cai, Siqi and Xie, Longhan and Li, Haizhou and Schultz, Tanja},
  journal={IEEE Transactions on Biomedical Engineering}, 
  title={STAnet: A Spatiotemporal Attention Network for Decoding Auditory Spatial Attention From EEG}, 
  year={2022},
  volume={69},
  number={7},
  pages={2233-2242},
}

@article{b33,
year = {2022},
month = {oct},
publisher = {IOP Publishing},
volume = {19},
number = {5},
pages = {056035},
author = {Jiang, Yifan and Chen, Ning and Jin, Jing},
title = {Detecting the locus of auditory attention based on the spectro-spatial-temporal analysis of EEG},
journal = {Journal of Neural Engineering}
}

@misc{b34,
      title={DGSD: Dynamical Graph Self-Distillation for EEG-Based Auditory Spatial Attention Detection}, 
      author={Cunhang Fan and Hongyu Zhang and Wei Huang and Jun Xue and Jianhua Tao and Jiangyan Yi and Zhao Lv and Xiaopei Wu},
      year={2023},
      eprint={2309.07147},
      archivePrefix={arXiv},
      primaryClass={eess.SP},
}

@article{b35,
  author={Biesmans, Wouter and Das, Neetha and Francart, Tom and Bertrand, Alexander},
  journal={IEEE Transactions on Neural Systems and Rehabilitation Engineering}, 
  title={Auditory-Inspired Speech Envelope Extraction Methods for Improved {EEG}-Based Auditory Attention Detection in a Cocktail Party Scenario}, 
  year={2017},
  volume={25},
  number={5},
  pages={402-412},
}

@article{b36,
  title={Noise-robust cortical tracking of attended speech in real-world acoustic scenes},
  author={Fuglsang, S{\o}ren A and Dau, Torsten and Hjortkj{\ae}r, Jens},
  journal={NeuroImage},
  volume={156},
  pages={435--444},
  year={2017},
  publisher={Elsevier},
}

@inproceedings{b37,
  title     = {DBPNet: Dual-Branch Parallel Network with Temporal-Frequency Fusion for Auditory Attention Detection},
  author    = {Ni, Qinke and Zhang, Hongyu and Fan, Cunhang and Pei, Shengbing and Zhou, Chang and Lv, Zhao},
  booktitle = {Proceedings of the Thirty-Third International Joint Conference on
               Artificial Intelligence, {IJCAI-24}},
  publisher = {International Joint Conferences on Artificial Intelligence Organization},
  editor    = {Kate Larson},
  pages     = {3115--3123},
  year      = {2024},
  month     = {8},
  note      = {Main Track},
}
\end{document}